\documentclass[graybox]{svmult}

\usepackage{type1cm}        % activate if the above 3 fonts are
\usepackage{makeidx}         % allows index generation
\usepackage{graphicx}        % standard LaTeX graphics tool
\usepackage{multicol}        % used for the two-column index
\usepackage[bottom]{footmisc}% places footnotes at page bottom

\usepackage{bbold} %for the identity matrix

\usepackage{newtxtext}       % 
\usepackage{newtxmath}       % selects Times Roman as basic font
\usepackage[colorlinks=true,citecolor=blue,linkcolor=blue,urlcolor=black]{hyperref}
\usepackage{bbm,dsfont}

\def\id{{\mathds 1}}
\newcommand{\ket}[1]{\left\vert#1\right\rangle}
\newcommand{\bra}[1]{\left\langle#1\right\vert}

\newcommand{\average}[1]{\left\langle#1\right\rangle}

\newcommand{\oper}[2]{\left\vert#1\rangle\!\langle#2\right\vert}

\renewcommand{\S}{{{}_S}}
\newcommand{\R}{{{}_R}}

\newcommand{\oS}{{{}_{\overline{S}}}}
\newcommand{\oR}{{{}_{\overline{R}}}}

\makeindex             % used for the subject index
\begin{document}

\title*{Quantum map approach to entanglement transfer and generation in spin chains}
\author{S. Lorenzo, F. Plastina, M. Consiglio, T. J. G. Apollaro}
\institute{S. Lorenzo \at dipartimento di Fisica e Chimica, Universit\'a di Palermo \email{salvatore.lorenzo@unipa.it}
\and F. Plastina \at dip. Fisica, Universit\'a della Calabria, 87036 Arcavacata di Rende (CS), Italy \email{francesco.plastina@fis.unical.it}
\and M. Consiglio \at Department of Physics, University of Malta, Msida MSD 2080, Malta \email{mirko.consiglio.16@um.edu.mt}
\and T. J. G. Apollaro \at Department of Physics, University of Malta, Msida MSD 2080, Malta \email{tony.apolalro@um.edu.mt}}
\maketitle

\abstract{Quantum information processing protocols are efficiently implemented on spin-$\frac{1}{2}$ networks. A quantum communication protocol generally involves a certain number of parties having local access to a subset of a larger system, whose intrinsic dynamics are exploited in order to perform a specific task. In this chapter, we address such a scenario with the quantum dynamical map formalism, where the dynamics of the larger system is expressed as a quantum map acting on the parties' access to their respective subsets of spins. We reformulate widely investigated protocols, such as one-qubit quantum state transfer and two-qubit entanglement distribution, with the quantum map formalism and demonstrate its usefulness in exploring less investigated protocols such as multi-qubit entanglement generation.}

\section{Introduction}
\label{sec:1}
Due to their formal analogy to quantum registers, quantum spin-$\frac{1}{2}$ networks have become the ideal testbed for many quantum information processing (QIP) protocols, ranging from quantum key distribution to quantum computation~\cite{Nielsen2010}. The availability of accurate theoretical models governing their dynamics, being amenable to solutions through either analytical techniques (especially for one-dimensional systems~\cite{Franchini2017}), or powerful numerical techniques, such as those based on tensor networks algorithms~\cite{Orus2019}, allows for the investigation of various and distinct protocols. These include, on the one hand, standard QIP protocols such as one-qubit quantum state transfer or bipartite Bell-type entanglement generation, taking place in an Hilbert space having dimensionality higher than that generally accessible via exact diagonalisation techniques. On the other hand, new QIP protocols are being introduced, aimed at exploring the complexity of the geometry of high-dimensional Hilbert spaces~\cite{Society2004}, such as in particular quantum state transfer \cite{Qin2013}. At the same time, remarkable progress has been made in order to experimentally verify these QIP protocols and communication processes on a variety of experimental platforms, with which the simulation of some quantum spin networks has been successfully achieved, using: cold atoms~\cite{Jaksch2005, Simon2011,Bloch2012}, Rydberg atoms~\cite{Labuhn2016b, PhysRevX.8.021069,Browaeys2020}, integrated photonic chips~\cite{Pitsios2017a}, trapped ions~\cite{Porras2004, Friedenauer2008,Monroe2021}, atom-cavity systems~\cite{Kay2008, Noh2017}, and superconducting circuits~\cite{Heras2014,Vepsalainen2020}, among others.
%%%%%%%%%%%%%%%%%%%%%%%%%%%%%%%%%%%%%%%%%%%%%%%%%%%%%%%%%%%%%%%%%%%%%%%%%%%%%%%%%%%%%%%%%%%%%%%%%%%%%%

\begin{figure}
	\centering
	\includegraphics[width=0.8\linewidth]{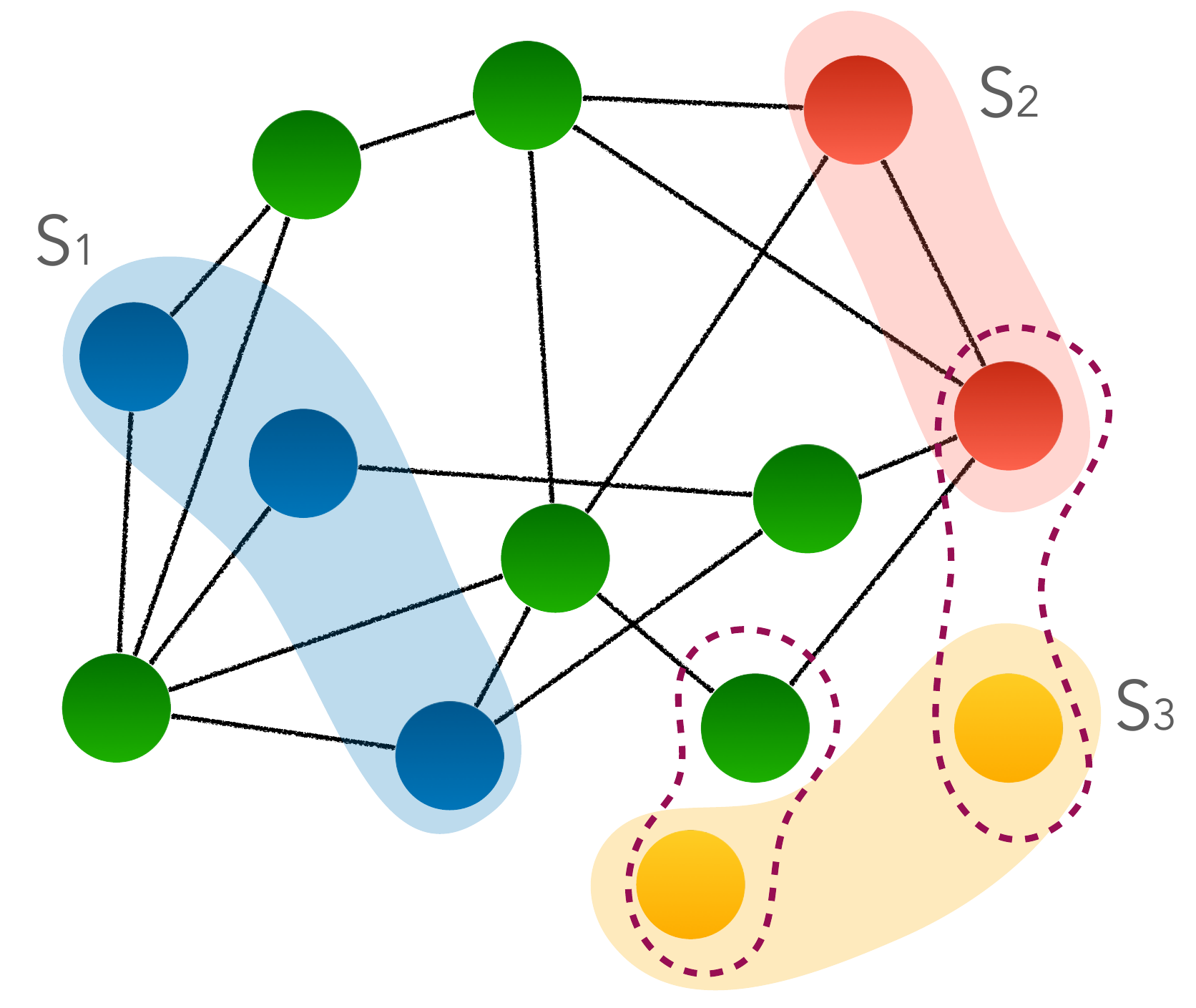}
	\caption{Sketch of a generic quantum spin-$\frac{1}{2}$ network $S$, where each party $P_i$ has access to the subset $S_i$ ($i=1,2,3$), on which local quantum  operations are allowed. In addition, the parties can exchange classical communication among them. The shaded area encloses the subsets $S_i$, the dashed lines indicate quantum correlations between the spins entering the QIP protocol, and the black continuous lines are the interactions among the spins in the network.}
	\label{fig:sketch}
\end{figure}

In the realm of the QIP tasks implementable with spin systems, a series of basic operations have been identified falling into the class of quantum communication protocols~\cite{Gisin2007}, which includes both the distribution and the generation of quantum resources at different space-time locations. A common communication scenario, depicted in Fig.~\ref{fig:sketch}, is represented by the circumstance in which a certain number of parties $P_i, i = 1, \ldots n$, each one having access to only a relatively small subset $S_i$ of a larger physical system $S$ (e.g., to a limited number of sites of a spin network), is required to receive/transfer quantum information from/to the others. Each party  is then  allowed to perform only {\it local quantum operations}; that is, $P_i$ is able to act on $S_i$ only, with the complementary system, $\Bar{S}_i:S\setminus S_i$, being inaccessible to any quantum operation it can carry out. Additionally, one can also allow for classical communication; i.e., the exchange of classical information among the parties. This combination is referred to as LOCC (local operations and classical communication) and the properties of LOCC operations determine to a large extent the fundamental limits for the performance of QIP protocols~\cite{Pirandola2017}. 
%%%%%%%%%%%%%%%%%%%%%%%%%%%%%%%%%%%%%%%%%%%%%%%%%%%%%%%%%%%%%%%%%%%%%%%%%%%%%%%%%%%%%%%%%%%%%%%%%%%%%%

In this chapter, we will employ the quantum dynamical map formalism, typically used in the theory of open quantum systems~\cite{book:19074}, to illustrate QIP protocols for quantum state transfer, entanglement distribution and generation on a homogeneous system, made up of a spin-$\frac{1}{2}$ network. We will assume that each party $i$ has access to a subset of $n_i$ spins of the network and has control over the interactions of the spins of this subset with the complementary system. Our aim is to derive the form of the dynamical map and, whenever possible, its analytical expression, in order to determine which LOCC operations maximise the efficiency of the investigated QIP protocols. We will focus, in particular, on the case of quantum dynamical maps obtained from spin Hamiltonians exhibiting $U(1)$ symmetry and, in order to obtain analytical results, investigate specific instances where the spin Hamiltonian is integrable.

%%%%%%%%%%%%%%%%%%%%%%%%%%%%%%%%%%%%%%%%%%%%%%%%%%%%%%%%%%%%%%%%%%%%%%%%%%%%%%%%%%%%%%%%%%%%%%%%%%%%%

The chapter is organised as follows: in Sec.~\ref{sec:1map}, we review the quantum dynamical map formalism, and apply it to $U(1)$-symmetric Hamiltonians in Sec.~\ref{sec:Ham}. In Sec.~\ref{sec:1q}, we illustrate the use of the formalism for case of single qubit quantum state transfer and Bell-state distribution; in Sec.~\ref{sec:2q}, we derive the two-qubit map for entanglement generation and distribution; in Sec.~\ref{sec:4q}, we explore the use of a 4-qubit dynamical map for investigating multipartite entanglement; and, finally, in Sec.~\ref{sec:conc} we draw our conclusions and provide some outlooks.

\section{Quantum dynamical maps}\label{sec:1map}

A quantum dynamical map between two systems associated to the Hilbert spaces $\mathcal{H}_1$, $\mathcal{H}_2$ can be identified with a linear homomorphism $\Phi:\mathcal{D}(\mathcal{H}_1){\rightarrow}\mathcal{D}(\mathcal{H}_2)$ mapping the space of density matrices acting on the {\it input} Hilbert space, into the space of density matrices acting on the {\it output} Hilbert space. Therefore, any $\Phi$ preserves the basic properties of the quantum states:
\begin{itemize}
    \item Self-adjointness: $\Phi(\rho^\dagger)^\dagger=\Phi(\rho)$,
    \item Complete positivity: $\Phi(\rho)>0$,
    \item Normalisation condition: $\text{Tr}(\rho)=1$,
    \item Linearity: $\Phi(a\rho_A+b\rho_B)=a\Phi(\rho_A)+b\Phi(\rho_B)$.
\end{itemize}
For the purpose of our investigation, we consider, from now on, finite-dimensional Hilbert spaces. For such finite dimensional case, the space of linear maps $\mathcal{L}(\mathbb{C}^n)$ can be identified with the algebra of $n\times n$ complex matrices, $M_n$. Any orthonormal basis $\ket{i}:i{=}1,...,n$ in $\mathbb{C}^n$ allows to define the orthonormal basis of elementary matrices in $M_n$:
\begin{eqnarray}
e_\alpha=e_{ij}=\oper{i}{j} \,,
\end{eqnarray}
and any map can be expressed as
\begin{eqnarray}
\Phi(\rho)=\sum_{\alpha,\beta}A_\alpha^\beta \text{Tr}\{e_\beta\rho\} e_\alpha \,.
\end{eqnarray}
In this basis, the matrix $A$ satisfies the following properties:
\begin{eqnarray}
    A_\alpha^\beta\equiv A_{ij}^{nm} \,, \qquad\qquad
    \begin{cases}\left ( {A}_{ij}^{nm} \right)^*=A_{ji}^{mn} \,, \\
    A_{ii}^{nm}=\delta_{mn} \,, \\
    A_{ij}^{nm}\rho_{nm}>0 \,.
    \end{cases}
\end{eqnarray}
The map $\Phi$ is completely positive if there exists a family of $N$ operators $K_i:i{=}1,...,N$ in $M_n$, which satisfy the condition $\sum_i K^{\dag}_i K_i = \id$, and such that $\Phi$ can be decomposed as \cite{Stinespring, Kraus, Zyczkowskibook}
\begin{eqnarray}
\label{Kraus}
\Phi\left(\rho\right)=\sum_{k=1}^N E_k\rho E_k^\dagger \,, 
\end{eqnarray}
By explicitly writing the matrix elements, we immediately find the relationship between map and Kraus operators:
\begin{eqnarray}
\left(\Phi\left(\rho\right)\right)_{ij}=\sum_{k=1}^N \sum_{nm}\left(E_k\right)_{in}\left(\rho\right)_{nm}\left(E_k^\dagger\right)_{mj}=\sum_{nm}A_{ij}^{nm}\left(\rho\right)_{nm}\label{Kraus2Map}
\end{eqnarray}
Note that, from Eq. \eqref{Kraus}, it is evident that for $N=1$ the map $\Phi$ represents a unitary map $\mathcal{U}=U\rho U^\dagger$. 

In this chapter, we are interested in entanglement generation and transfer between two sub-parties, that we dub sender and receiver, each of them taking care, controlling and possibly making measurements on a subset of the system's spins. We denote the states of the subsystems pertaining to sender and receiver as $\rho_\S$ and $\rho_\R$, respectively, and assume that these are the marginals obtained from the state of a larger system $\sigma$, whose time evolution is dictated by a Hamiltonian generating a unitary map, i.e. $\sigma(t) = \mathcal{U}(t)\left(\sigma(0)\right)$. We assume that the initial state of this larger system, $\sigma(0)$, is a product state between $\rho_\S$ and a reference pure state $\ket{\Psi}_\oS$. In other words, we are concerned with maps of the following form
\begin{eqnarray}
\label{eq_map_U}
\rho_\R=\Phi(\rho_\S)=\text{Tr}_{\oR}\{\mathcal{U}(\rho_\S\otimes\ket{\Psi}_\oS\!\bra{\Psi})\}
\end{eqnarray}
where we indicate with $\text{Tr}_\oR$ the partial trace over all but $\text{R}$ degrees of freedom. Denoting with $\ket{r}_\oR$ an orthonormal basis of $\mathcal{H}_\oR$ we have 
\begin{eqnarray}
\label{Kraus2}
\rho_\R=\Phi(\rho_\S)=\sum_{r}\left(_\oR\!\bra{r}U\ket{\Psi}_\oS\right)\rho_\S\left( _\oS\!\bra{\Psi}U^\dagger\ket{r}_\oR\right)=\sum_{r}E_r\rho_\S E_r^\dagger \,.
\end{eqnarray}
It is important to remember that the Kraus operators $E_r$ are time dependent and that they depend on the choice of the basis in $\oR$, and, more generally, on the choice of sender and receiver subsets themselves.
The matrix representation of the corresponding map can be found according to Eq. \eqref{Kraus2Map}.

\section{\textit{U}(1)-symmetric Hamiltonians}\label{sec:Ham}

In this Section we will derive the general form of the quantum dynamical map in Eq.~\eqref{Kraus2Map} when the unitary evolution operator entering Eq.~\eqref{eq_map_U} exhibits $U(1)$ symmetry. Without loss of generality we focus on spin-$\frac{1}{2}$ Hamiltonians with isotropic Heisenberg-type interactions in the $XY$-plane:
\begin{align}
    \label{eq:gen_Ham}
    \hat{H}=\sum_{i,j}\left(J_{ij}\left(\hat{\sigma}_i^x\hat{\sigma}_j^x+\hat{\sigma}_i^y\hat{\sigma}_j^y\right)+\Delta_{ij}\hat{\sigma}_i^z\hat{\sigma}_j^z\right)+\sum_i h_i \hat{\sigma}_j^z~,
\end{align}
where $\hat{\sigma}_i^{\alpha}$ ($\alpha=x,y,z$) are the usual Pauli matrices, $i$ denotes the site-index, $J_{ij}$ and $\Delta_{ij}$ are, respectively, the two-qubit interaction terms in the $XY$-plane and along the $Z$-axis, and $h_i$ is the magnetic field along the $Z$-axis. In fact, the class of Hamiltonians exhibiting the $U(1)$ symmetry is larger than that described by Eq.~\eqref{eq:gen_Ham}, and encompasses Hamiltonians with Dzyaloshinskii–Moriya~\cite{Verma2017b} and $XY$-isotropic cluster interaction terms~\cite{Campbell2013d}, among others. 

In terms of spin operators, the $U(1)$ symmetry implies that the total magnetisation along the $Z$-axis, $\average{\hat{M}}=\sum_{i=1}^N\average{\hat{\sigma}^z_i}$, is a conserved quantity and the  operator $\hat{M}$ commutes with $U$. Hence, it is possible to divide the whole Hilbert space into invariant subspaces, labelled by the eigenvalues of $\hat{M}$, with each subspace having the dimension determined by the degeneracy of the eigenvalue, $\binom{N}{i}$, where $i$ denotes the number of flipped spins. Indeed, by writing the spectral decomposition of $\hat{M}$ as \begin{eqnarray}
\hat{M}= \sum_k\sum_d \lambda_k \ket{\phi_k^d}\bra{\phi_k^d} \,,
\end{eqnarray}
we know that $U\ket{\phi_k^d}$ is an eigenstate of $\hat{M}$ with eigenvalue $\lambda_k$, i.e.
\begin{eqnarray}
\bra{\phi_{k'}^{d'}}U\ket{\phi_k^d}=\bra{\phi_{k'}^{d'}}U\ket{\phi_k^d}\delta_{kk'}=(f_k)^{d'}_{d}\delta_{kk'} \,.
\end{eqnarray}
Thus, we can then write $U$ as a direct sum of unitary operators acting in each subspace
\begin{eqnarray}
U=U_0 \oplus U_1 \oplus U_2 \oplus ...
\end{eqnarray}
If we now observe the elements of the Kraus operators entering in Eq. \eqref{Kraus2}, they take the form 
\begin{eqnarray}\label{eq:Mag_K}
(E_r)_{in}=\;_\R\!\bra{i}E_r\ket{n}_\S=\;_\R\!\bra{i}\;\!_\oR\!\bra{r}U\ket{\Psi}_\oS\ket{n}_\S \,.
\end{eqnarray}
This inherently places some constraints on the elements of the Kraus operators, and consequently on the map elements. Indeed, if the state $\ket{\Psi}_\oS\ket{n}_\S$ belongs in a given subspace, then $_\R\!\bra{i}_\oR\!\bra{r}$ must belong to that subspace as well in order for the above equation to be non-zero. Without loss of generality, we take $\ket{\Psi}_\oS\ket{n}_\S$ as living in the $n$-th supspace, implying that $i+r=n$. 
In the following, we will assume $\ket{\Psi}_\oS=\ket{\mathbf{0}}$, i.e., a fully polarised state.

%%%%%%%%%%%%%%%%%%%%%%%%%%%%%%%%%%%%%%%%%%%%%%%%%%%%%%%%%%%%%%%%%%%%%%%%%%%%%%%%%%%%%%%%%%%%%%%%%
\section{One-qubit map}
\label{sec:1q}

To begin our analysis, let us consider the simplest case: a map from qubit $i$ (the sender) to qubit $j$ (the receiver): $\hat{\rho}_j(t)=\Phi(t)\hat{\rho}_i(0)$, where $i$ and $j$ are (possibly identical) positions in a spin network. In this case, we have two possible values of $r=0,1$ and, consequently, two Kraus operators:
\begin{eqnarray}
E_0=\begin{pmatrix}
1 &0\\
0&f_i^j
\end{pmatrix} \,, \qquad 
E_1^k=\begin{pmatrix}
0 &0\\
f_i^k&0
\end{pmatrix} \; \text{with }k\neq j\,.
\end{eqnarray} 
By using Eq. \eqref{Kraus2Map}, we can write the map as
\begin{align}\label{E_1ex_map}
\begin{pmatrix}
\rho_{00}\\
\rho_{01}\\
\rho_{10}\\
\rho_{11}\\
\end{pmatrix}_j=
\begin{pmatrix}
1 &0&0&1-\left|f_i^j\right|^2\\
0&f_i^j&0&0\\
0&0&\left(f_i^j\right)^*&0\\
0&0&0&\left|f_i^j\right|^2
\end{pmatrix}
\begin{pmatrix}
\rho_{00}\\
\rho_{01}\\
\rho_{10}\\
\rho_{11}\\
\end{pmatrix}_i \,,
\end{align}
where we use the completeness relation $\sum_n \left|f_i^n\right|^2=1$.\\
In this simple case, perfect state transfer, i.e. $\hat{\rho}^{j}_{nm}=\hat{\rho}^{i}_{nm}$, entails $f_i^j=\left(f_i^j\right)^*=1$. A considerable amount of research has been performed in order to investigate the conditions that allow to maximise the transition amplitude~\cite{PhysRevA.93.012343}.

\begin{figure}
	\centering
	\includegraphics[width=0.9\linewidth]{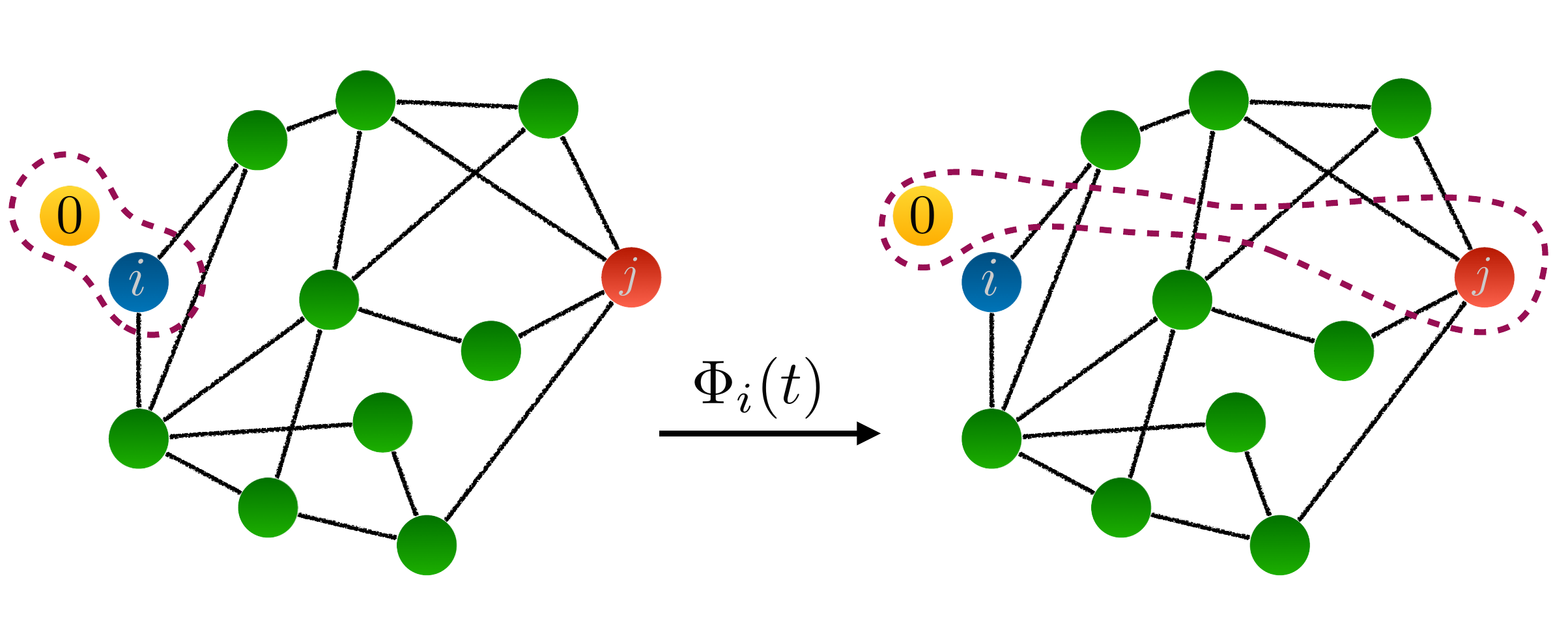}
	\caption{A schematic picture of an entanglement distribution protocol. Initially, the external qubit $0$ is entangled with qubit $i$ and the aim is to exploit the map in Eq.~\ref{E_map_2Bose} to entangle the former with qubit $j$.}
	\label{fig:dist_ent_1}
\end{figure}

In Bose's original protocol~\cite{Bose2003} this is achieved by a local magnetic field acting on the spins. The map in Eq.~\ref{E_1ex_map} is also informative about remote state preparation protocols: the coherence (in the computational basis) of spin $j$ cannot increase with respect to that of spin $i$ under the action of this map as the off-diagonal elements of the output density matrix can only be suppressed, or, at most, maintain their initial amplitude.

Apart from quantum state transfer protocols, the map in Eq.~\ref{E_1ex_map} can be used to analyse entanglement distribution protocols, like the one reported in Refs.~\cite{Bose2003,Bayat2011} and sketched in Fig. \ref{fig:dist_ent_1}. Explicitly, the map is given by 
\begin{align}
    \hat{\rho}(t)_{0j}=\left(\mathbbm{1}_0\otimes \Phi_i(t)\right)\hat{\rho}(0)_{0i}~,
\end{align}
where $\mathbbm{1}_0$ is the $4$-dimensional identity map acting on qubit $0$ and $\Phi_i(t)$ is given by the map in Eq.~\ref{E_1ex_map} acting on qubit $i$.
The map reads
\begin{eqnarray}
\label{E_map_2Bose}
\begin{small}
 \begin{pmatrix}
\rho_{00}\\
\rho_{01}\\
\rho_{02}\\
\rho_{03}\\
\rho_{10}\\
\rho_{11}\\
\rho_{12}\\
\rho_{13}\\
\rho_{20}\\
\rho_{21}\\
\rho_{22}\\
\rho_{23}\\
\rho_{30}\\
\rho_{31}\\
\rho_{32}\\
\rho_{33}\\
\end{pmatrix}_{0,j}=   \left(
\begin{array}{cccccccccccccccc}
 1 & 0 & 0 & 0 & 0 & 1{-}| f| ^2 & 0 & 0 & 0 & 0 & 0 & 0 & 0 & 0 & 0 & 0 \\
 0 & f & 0 & 0 & 0 & 0 & 0 & 0 & 0 & 0 & 0 & 0 & 0 & 0 & 0 & 0 \\
 0 & 0 & 1 & 0 & 0 & 0 & 0 & 1{-}| f| ^2 & 0 & 0 & 0 & 0 & 0 & 0 & 0 & 0 \\
 0 & 0 & 0 & f & 0 & 0 & 0 & 0 & 0 & 0 & 0 & 0 & 0 & 0 & 0 & 0 \\
 0 & 0 & 0 & 0 & f^* & 0 & 0 & 0 & 0 & 0 & 0 & 0 & 0 & 0 & 0 & 0 \\
 0 & 0 & 0 & 0 & 0 & | f| ^2 & 0 & 0 & 0 & 0 & 0 & 0 & 0 & 0 & 0 & 0 \\
 0 & 0 & 0 & 0 & 0 & 0 & f^* & 0 & 0 & 0 & 0 & 0 & 0 & 0 & 0 & 0 \\
 0 & 0 & 0 & 0 & 0 & 0 & 0 & | f| ^2 & 0 & 0 & 0 & 0 & 0 & 0 & 0 & 0 \\
 0 & 0 & 0 & 0 & 0 & 0 & 0 & 0 & 1 & 0 & 0 & 0 & 0 & 1{-}| f| ^2 & 0 & 0 \\
 0 & 0 & 0 & 0 & 0 & 0 & 0 & 0 & 0 & f & 0 & 0 & 0 & 0 & 0 & 0 \\
 0 & 0 & 0 & 0 & 0 & 0 & 0 & 0 & 0 & 0 & 1 & 0 & 0 & 0 & 0 & 1{-}| f| ^2 \\
 0 & 0 & 0 & 0 & 0 & 0 & 0 & 0 & 0 & 0 & 0 & f & 0 & 0 & 0 & 0 \\
 0 & 0 & 0 & 0 & 0 & 0 & 0 & 0 & 0 & 0 & 0 & 0 & f^* & 0 & 0 & 0 \\
 0 & 0 & 0 & 0 & 0 & 0 & 0 & 0 & 0 & 0 & 0 & 0 & 0 & | f| ^2 & 0 & 0 \\
 0 & 0 & 0 & 0 & 0 & 0 & 0 & 0 & 0 & 0 & 0 & 0 & 0 & 0 & f^* & 0 \\
 0 & 0 & 0 & 0 & 0 & 0 & 0 & 0 & 0 & 0 & 0 & 0 & 0 & 0 & 0 & | f| ^2 \\
\end{array}
\right)
\begin{pmatrix}
\rho_{00}\\
\rho_{01}\\
\rho_{02}\\
\rho_{03}\\
\rho_{10}\\
\rho_{11}\\
\rho_{12}\\
\rho_{13}\\
\rho_{20}\\
\rho_{21}\\
\rho_{22}\\
\rho_{23}\\
\rho_{30}\\
\rho_{31}\\
\rho_{32}\\
\rho_{33}\\
\end{pmatrix}_{0,i}
\end{small}
\end{eqnarray}

\begin{figure}
	\centering
	\includegraphics[width=0.8\linewidth]{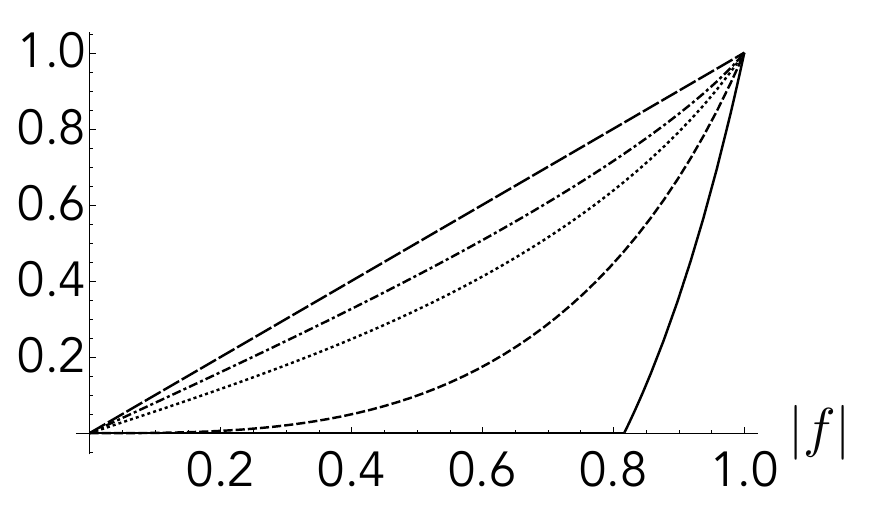}
	\caption{Ratio of transferred entanglement vs $\left|f\right|$ for a Werner state for different values of $p$ in Eq.~\ref{eq:Werner}. The curves, from bottom to top, are drawn for $p=0.4,0.5,0.7,0.9,1$.}
	\label{fig:dist_ent_W}
\end{figure}

In the original entanglement distribution protocol given in Ref.~\cite{Bose2003}, the entanglement encoded in a singlet state on sites $0$ and $i$ is distributed to sites $0$ and $j$ resulting in a Concurrence $C=\left|f_i^j(t)\right|$. From the map in Eq.~\ref{E_map_2Bose}, it is evident that the same holds true for any Bell state. On the other hand, for $X$-type density matrices~\cite{Amico2004b}, the distributed entanglement does not increase linearly with the (modulus) of the transition amplitude $f$, but instead reads
\begin{align}
\label{eq:ap_p}
    C=2\max\left[0, C_1,C_2\right]~,
\end{align}
where
\begin{align}
    C_1=\left|f\right|\left(\left|\rho_{12}\right|-\sqrt{\rho_{33}\left(\rho_{00}+\rho_{11}\left(1-\left|f\right|^2\right)\right)}\right) \,, \\
    C_2=\left|f\right|\left(\left|\rho_{03}\right|-\sqrt{\rho_{11}\left(\rho_{22}+\rho_{33}\left(1-\left|f\right|^2\right)\right)}\right) \,,
\end{align}
denote the so called anti-parallel and parallel concurrences, respectively~\cite{Fubini2006a}. Looking carefully at the map, one can see that the ratio of transferred entanglement over initial entanglement depends only on the transition amplitude, and not on the fact that entanglement is of the parallel or anti-parallel type. The dependence on $\left|f \right|$ is not linear, as in the pure Bell-states scenario, and in Fig.~\ref{fig:dist_ent_W} we show the ratio of the transferred concurrence vs the initial concurrence ($C = \frac{3p-1}{2}$) for an initial Werner State 
\begin{align}
\label{eq:Werner}
\hat{\rho}_W=p\ket{\Psi_B}\bra{\Psi_B}+\left(1-p\right)\frac{\mathbbm{1}}{4}~,
\end{align}
where $\ket{\Psi_B}$ is any Bell state.

\begin{figure}
	\centering
	\includegraphics[width=0.9\linewidth]{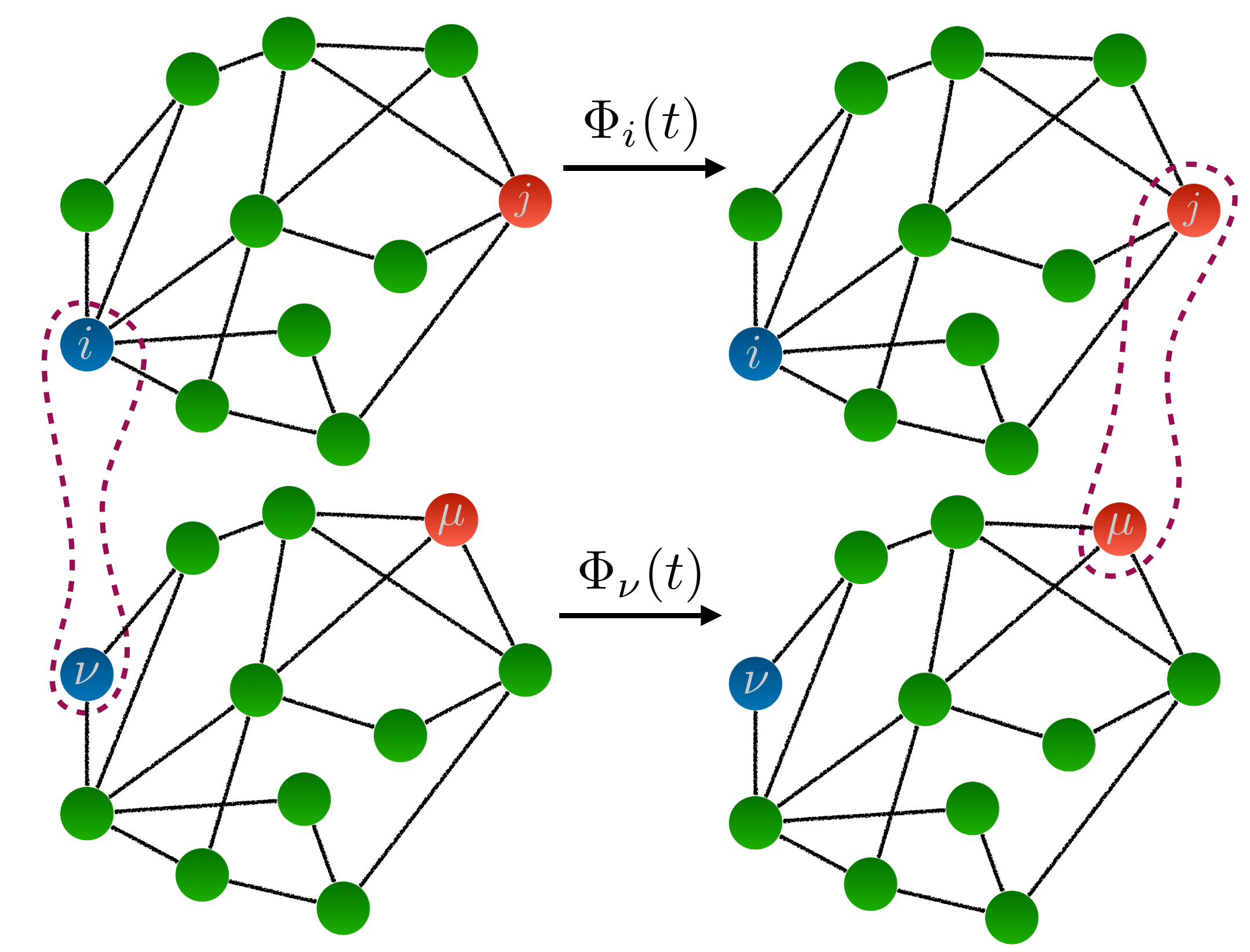}
	\caption{A schematic picture of a entanglement distribution protocol. Initially qubit $i$ is entangled with qubit $\nu$, and the aim is to exploit the map in Eq.~\ref{E_map_2Bose} to entangle qubit $j$ with qubit $\mu$.}
	\label{fig:dist_ent_2}
\end{figure}

Now we consider the case where the entanglement distribution protocol is designed in order to send to sites $\left(n,\mu\right)$ (the receiver sites) the entanglement initially shared between sites $\left(i,\nu\right)$ (sending sites) using two independent spin networks. A particular instance of this setup is given in Fig.~\ref{fig:dist_ent_2}, and both there and in the setting of the problem above, Latin (Greek) letters are used to denote the sites on the first (second) chain.  This protocol is reminiscent of the dual rail encoding protocol for sending a single qubit state~\cite{Bose2007}.

The map is given by 
\begin{align}
    \hat{\rho}(t)_{j\mu}=\left(\Phi_i(t)\otimes \Phi_{\nu}(t)\right)\hat{\rho}(0)_{i\nu}~,
\end{align}
and reads 
\begin{small}
\begin{align}
\left(
\begin{array}{cccccccccccccccc}
 1 & 0 & 0 & 0 & 0 & 1{-}| g| ^2 & 0 & 0 & 0 & 0 & 1{-}| f| ^2 & 0 & 0 & 0 & 0 & \left(1{-}| f| ^2\right) \left(1{-}| g| ^2\right) \\
 0 & g & 0 & 0 & 0 & 0 & 0 & 0 & 0 & 0 & 0 & g \left(1{-}| f| ^2\right) & 0 & 0 & 0 & 0 \\
 0 & 0 & f & 0 & 0 & 0 & 0 & f \left(1{-}| g| ^2\right) & 0 & 0 & 0 & 0 & 0 & 0 & 0 & 0 \\
 0 & 0 & 0 & f g & 0 & 0 & 0 & 0 & 0 & 0 & 0 & 0 & 0 & 0 & 0 & 0 \\
 0 & 0 & 0 & 0 & g^* & 0 & 0 & 0 & 0 & 0 & 0 & 0 & 0 & 0 & \left(1{-}| f| ^2\right) g^* & 0 \\
 0 & 0 & 0 & 0 & 0 & | g| ^2 & 0 & 0 & 0 & 0 & 0 & 0 & 0 & 0 & 0 & \left(1{-}| f| ^2\right) | g| ^2 \\
 0 & 0 & 0 & 0 & 0 & 0 & f g^* & 0 & 0 & 0 & 0 & 0 & 0 & 0 & 0 & 0 \\
 0 & 0 & 0 & 0 & 0 & 0 & 0 & f | g| ^2 & 0 & 0 & 0 & 0 & 0 & 0 & 0 & 0 \\
 0 & 0 & 0 & 0 & 0 & 0 & 0 & 0 & f^* & 0 & 0 & 0 & 0 & \left(1{-}| g| ^2\right) f^* & 0 & 0 \\
 0 & 0 & 0 & 0 & 0 & 0 & 0 & 0 & 0 & g f^* & 0 & 0 & 0 & 0 & 0 & 0 \\
 0 & 0 & 0 & 0 & 0 & 0 & 0 & 0 & 0 & 0 & | f| ^2 & 0 & 0 & 0 & 0 & | f| ^2 \left(1{-}| g| ^2\right) \\
 0 & 0 & 0 & 0 & 0 & 0 & 0 & 0 & 0 & 0 & 0 & g | f| ^2 & 0 & 0 & 0 & 0 \\
 0 & 0 & 0 & 0 & 0 & 0 & 0 & 0 & 0 & 0 & 0 & 0 & f^* g^* & 0 & 0 & 0 \\
 0 & 0 & 0 & 0 & 0 & 0 & 0 & 0 & 0 & 0 & 0 & 0 & 0 & | g| ^2 f^* & 0 & 0 \\
 0 & 0 & 0 & 0 & 0 & 0 & 0 & 0 & 0 & 0 & 0 & 0 & 0 & 0 & | f| ^2 g^* & 0 \\
 0 & 0 & 0 & 0 & 0 & 0 & 0 & 0 & 0 & 0 & 0 & 0 & 0 & 0 & 0 & | f| ^2 | g| ^2 \\
\end{array}
\right)~,
\end{align}
\end{small}
where we have used the short-hand notation $f=f_i^j$ and $g=f_{\nu}^{\mu}$.
If we assume that the two spin networks are identical and that the locations of the initially correlated sites and of the receiving ones are also the same on the two networks, the matrix above simplifies as $f=g$. Again, this is the same settings adopted in the dual-rail protocol~\cite{Burgarth2005} to attain perfect state transfer. In such a case, with $\hat{\rho}_{i\nu}$ being an $X$-type state, the anti-parallel and parallel concurrences reported in Eq.~\ref{eq:ap_p} are given by
\begin{align}
    \label{eq:dual_rail}
    &C_1=\left|f\right|^2\left(\left|\rho_{12}\right|-\sqrt{\rho_{33}\left(\rho_{00}+\left(1- \left|f\right|^2\right)\left(\rho_{11}+\rho_{22}+\left(1- \left|f\right|^2\right)\rho_{33} \right)\right)}\right)\\
    &C_2=\left|f\right|^2\left(\left|\rho_{03}\right|-\sqrt{\left(\rho_{11}+\left(1-\left|f\right|^2\right)\rho_{33}\right)\left(\rho_{22}+\left(1-\left|f\right|^2\right)\rho_{33}\right)}\right)~.
\end{align}

Due to the use of two channels, as depicted in Fig.~\ref{fig:dist_ent_2}, the ratio of transferred over initial entanglement depends, this time, not only on the transition amplitude, but also on the type of entanglement (whether parallel or anti-parallel). This is also the case when investigating the effect of the spin environment on the entanglement, which has been carried out in Ref.~\cite{Apollaro2010a}, by analysing the properties of the map $\hat{\rho}(t)_{j\mu}=\left(\Phi_i(t)\otimes \Phi_{\nu}(t)\right)\hat{\rho}(0)_{j\mu}$.

\begin{figure}
	\centering
	\includegraphics[width=\linewidth]{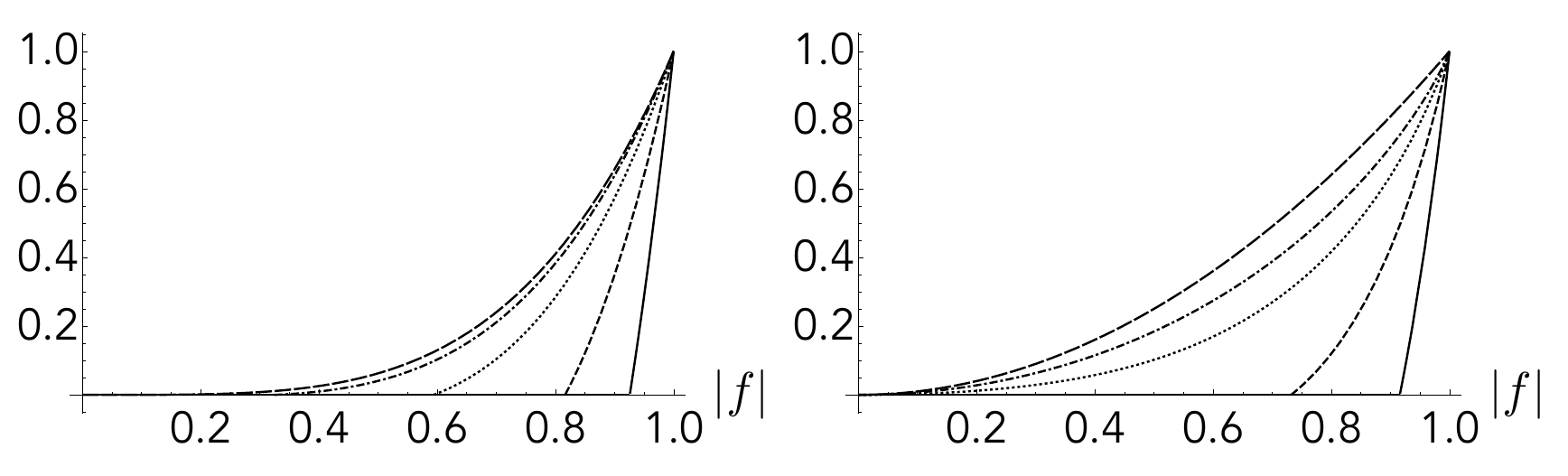}
	\caption{Ratio of transferred over initial entanglement vs. $\left|f\right|$ for a Werner state for different values of $p$ in Eq.~\ref{eq:Werner}. The curves, from bottom to top, are drawn for $p=0.4,0.5,0.7,0.9,1$. The left plot corresponds to a Werner state with maximally entangled component given by $\ket{\Psi_B}=\ket{\Phi^+}$; whereas, for the right plot we have chosen $\ket{\Psi_B}=\ket{\Psi^+}$}.
	\label{fig:dist_ent_B2}
\end{figure}

As one can expect, the anti-parallel entanglement is one, so that $C_1$ attains larger values with respect to $C_2$. Intuitively, this is due to the fact that in the former case, only one excitation is present in the system; whereas, in the latter, two excitations enter the dynamics. This leads to an increase in the effects of decoherence due to the dispersion of the extra excitation all over the network. A figure of merit describing the amount of transferred entanglement is reported in Fig.~\ref{fig:dist_ent_B2}, in the case of initial Werner states (Eq.~\ref{eq:Werner}) 

\section{Two-qubit map}
\label{sec:2q}
In this section, we derive the expression of $\Phi$ for the case of two sender qubits located at arbitrary positions $i$ and $j$ in a spin network.
For a generic two-qubit receiver at sites $n,m$ is instructive to write the map in matrix form.

\begin{eqnarray}
\label{E_2ex_map}
\begin{small}
\begin{pmatrix}
\rho_{00}\\
\rho_{01}\\
\rho_{02}\\
\rho_{03}\\
\rho_{10}\\
\rho_{11}\\
\rho_{12}\\
\rho_{13}\\
\rho_{20}\\
\rho_{21}\\
\rho_{22}\\
\rho_{23}\\
\rho_{30}\\
\rho_{31}\\
\rho_{32}\\
\rho_{33}
\end{pmatrix}_{i,j}\!\!\!\!\!=
 \begin{pmatrix}
 A_{00}^{00}&0&0&0&0&A_{00}^{11}&A_{00}^{12}&0&0&A_{00}^{21}&A_{00}^{22}&0&0&0&0&A_{00}^{33}\\
0& A_{01}^{01} & A_{01}^{02} & 0 & 0 & 0 & 0 & A_{01}^{13} & 0 & 0 & 0 & A_{01}^{23} & 0 & 0 & 0 & 0 \\
0 & A_{02}^{01} & A_{02}^{02} & 0 & 0 & 0 & 0 & A_{02}^{13} & 0 & 0 & 0 & A_{02}^{23} & 0 & 0 & 0 & 0 \\
0& 0 & 0 & A_{03}^{03} & 0 & 0 & 0 & 0 & 0 & 0 & 0 & 0 & 0 & 0 & 0 & 0 \\
0& 0 & 0 & 0 & A_{10}^{10} &  0  & 0 & 0 & A_{10}^{20} & 0 & 0  & 0 &0  & A_{10}^{31} & A_{10}^{32} & 0 \\
0& 0 & 0 & 0 & 0 & A_{11}^{11} & A_{11}^{12} & 0 & 0 & A_{11}^{21} & A_{11}^{22} & 0 & 0 & 0 & 0 & A_{11}^{33} \\
0& 0 & 0 & 0 & 0 & A_{12}^{11} & A_{12}^{12} & 0 & 0 & A_{12}^{21} & A_{12}^{22} & 0 & 0 & 0 & 0 & A_{12}^{33} \\
0& 0 & 0 & 0 & 0 & 0 & 0 & A_{13}^{13} & 0 & 0 & 0 & A_{13}^{23} & 0 & 0 & 0 & 0 \\
0& 0 & 0 & 0 &  A_{20}^{10} & 0 & 0 &  0 & A_{20}^{20} & 0 & 0 & 0 & 0 & A_{20}^{31} & A_{20}^{32} & 0 \\
0& 0 & 0 & 0 & 0 & A_{21}^{11} & A_{21}^{12} & 0 & 0 & A_{21}^{21} & A_{21}^{22} & 0 & 0 & 0 & 0 &  A_{21}^{33} \\
0& 0 & 0 & 0 & 0 & A_{22}^{11} & A_{22}^{12} & 0 & 0 & A_{22}^{21} & A_{22}^{22} & 0 & 0 & 0 & 0 & A_{22}^{33} \\
0& 0 & 0 & 0 & 0 & 0 & 0 & A_{23}^{13} & 0 & 0 & 0 & A_{23}^{23} & 0 & 0 & 0 & 0\\
0& 0 & 0 & 0 & 0 & 0 & 0 & 0 & 0 & 0 & 0 & 0 & A_{30}^{30} & 0 & 0 &  0\\
0& 0 & 0 & 0 & 0 & 0 & 0 & 0 & 0 & 0 & 0 & 0 & 0 & A_{31}^{31} & A_{31}^{32} &0 \\
0& 0 & 0 & 0 & 0 & 0 & 0 & 0 & 0 & 0 & 0 & 0 & 0 & A_{32}^{31} & A_{32}^{32} & 0 \\
0& 0 & 0 & 0 & 0 & 0 & 0 & 0 & 0 & 0 & 0 & 0 & 0 & 0 & 0 & A_{33}^{33}
\end{pmatrix}
\begin{pmatrix}
\rho_{00}\\
\rho_{01}\\
\rho_{02}\\
\rho_{03}\\
\rho_{10}\\
\rho_{11}\\
\rho_{12}\\
\rho_{13}\\
\rho_{20}\\
\rho_{21}\\
\rho_{22}\\
\rho_{23}\\
\rho_{30}\\
\rho_{31}\\
\rho_{32}\\
\rho_{33}
\end{pmatrix}_{n,m}
\end{small}
 \end{eqnarray}

The map in Eq.~\ref{E_2ex_map} can be written in a more compact form defining the Kraus operators in all of magnetisation sectors, as illustrated in Sec.\ref{sec:Ham}, where Eq.~\ref{eq:Mag_K} has elements
%where
\begin{eqnarray}
A=E_0\otimes E_0^*+E_1\otimes E_1^*+E_2\otimes E_2^* \,,
\end{eqnarray}
and
\begin{equation}
E_0=\left(
\begin{array}{cccc}
 1 & 0 & 0 & 0 \\
 0 & f_{j}^{m} & f_{i}^{m} & 0 \\
 0 & f_{j}^{n} & f_{i}^{n} & 0 \\
 0 & 0 & 0 & f_{ij}^{nm}
\end{array}
\right) \,, \,
E_1^{k\neq i,j}=\left(
\begin{array}{cccc}
 0 & f_{j}^{k} & f_{i}^{k} & 0 \\
 0 & 0 & 0 & f_{ij}^{km} \\
 0 & 0 & 0 & f_{ij}^{nk} \\
 0 & 0 & 0 & 0 
\end{array}\right) \,, \,
E_2^{k,l \neq i,j}=\left(
\begin{array}{cccc}
 0 & 0 & 0 & f_{ij}^{kl} \\
 0 & 0 & 0 & 0 \\
 0 & 0 & 0 & 0 \\
 0 & 0 & 0 & 0
\end{array}
\right) \,.
\end{equation}

This explicit form of the map shows, for example, that $\rho_{03}(t) = A_{03}^{03}\rho_{03}(0)$. As a consequence (and analogously to the impossibility of amplifying coherence in the single-qubit case), the Bell states $\ket{\Phi^{\pm}}$ cannot be generated by this map, irrespective of the initial state of the qubits.
On the other hand, the Bell states $\ket{\Psi^{\pm}}$ can be generated by LOCC as the coherences $\rho_{12}$ can be build up starting from $\rho_{11}$, $\rho_{22}$ and/or $\rho_{33}$, i.e., by locally flipping the spins on the sender and/or receiver sites.

From Ref.~\cite{Apollaro2015}, we can borrow the following two-qubit map's elements for the case in which the receiver coincides with the sender (so that, in fact, we are evaluating how good is information storage at sites $i,j$), $\hat{\rho}_{ij}(t)=\Phi_t\hat{\rho}_{ij}(0)$
\begin{align}
\label{E_2map}
&A_{00}^{00}=1~,~A_{00}^{11}=1-\left|f_i^{i}\right|^2-\left|f_i^{j}\right|^2~,~A_{00}^{22}=1-\left|f_j^{i}\right|^2-\left|f_j^{j}\right|^2 \,, \nonumber \\
&A_{00}^{33}=1-\left|f_{ij}^{mi}\right|^2-\left|f_{ij}^{mj}\right|^2-\left|f_{ij}^{ij}\right|^2~,~A_{00}^{12}=-f_i^{i}\left(f_j^{i}\right)^*-f_i^{i}\left(f_j^{i}\right)^* \,, \nonumber \\
&A_{00}^{21}=-f_j^{i}\left(f_i^{i}\right)^*-f_j^{j}\left(f_i^{j}\right)^* \,, \nonumber \\
&A_{01}^{01}=\left(f_i^j\right)^*~,~A_{01}^{02}=\left(f_j^j\right)^*~,~A_{01}^{13}=f_i^m\left(f_{ij}^{mj}\right)^*~,~A_{01}^{23}=f_j^m\left(f_{ij}^{mj}\right)^* \,, \nonumber \\
&A_{02}^{01}=\left(f_{i}^{i}\right)^*~,~A_{02}^{02}=\left(f_{j}^{i}\right)^*~,~A_{02}^{13}=f_i^m\left(f_{ij}^{mi}\right)^*~,~A_{02}^{23}=f_j^m\left(f_{ij}^{mi}\right)^* \,, \nonumber \\
&A_{03}^{03}=\left(f_{ij}^{ij}\right)^* \,, \nonumber \\
&A_{11}^{11}=\left|f_i^j\right|^2~,~A_{11}^{12}=f_i^j\left(f_j^i\right)^*~,~A_{11}^{21}=f_j^j\left(f_i^j\right)^*~,~A_{11}^{22}=\left|f_j^j\right|^2~,~A_{11}^{33}=\left|f_{ij}^{mj}\right|^2 \,, \nonumber \\
&A_{12}^{11}=f_i^j\left(f_i^i\right)^*~,~A_{12}^{12}=f_i^j\left(f_j^i\right)^*~,~
A_{12}^{21}=f_j^j\left(f_i^i\right)^*~,~A_{12}^{22}=f_j^j\left(f_j^i\right)^*~,~
A_{12}^{33}=f_{ij}^{mj}\left(f_{ij}^{mi}\right)^* \,, \nonumber \\
&A_{13}^{13}=f_i^j\left(f_{ij}^{ij}\right)^*~,~A_{13}^{23}=f_j^j\left(f_{i}^{ij}\right)^* \,, \nonumber \\
&A_{22}^{33}=\left|f_{ij}^{mi}\right|^2~,~A_{22}^{11}=\left|f_{i}^{i}\right|^2~,~A_{22}^{22}=\left|f_{j}^{i}\right|^2~,~A_{22}^{12}=f_{i}^{i}\left(f_{j}^{i}\right)^*~,~A_{22}^{21}=f_{j}^{i}\left(f_{i}^{i}\right)^* \,, \nonumber \\
&A_{23}^{13}=f_{i}^{i}\left(f_{ij}^{ij}\right)^*~,~A_{23}^{23}=f_{j}^{i}\left(f_{ij}^{ij}\right)^*~,~A_{33}^{33}=\left|f_{ij}^{ij}\right|^2 \,,
\end{align}
where, whenever the index $m$ appears, a summation over all $m\neq S,R$ is intended.

For the general case describing the transfer of a given two qubit state from the sending pair $i,j$ to the receiving pair $n,m$,  $\hat{\rho}_{nm}(t)=\Phi_t\hat{\rho}_{ij}(0)$, the matrix elements read:
\begin{eqnarray}
&& A_{00}^{00}=1 \, , \, A_{00}^{11}=f_{j}^{k} \left(f_{j}^{k}\right){}^* \, , \, A_{00}^{12}=f_{j}^{k} \left(f_{i}^{k}\right){}^* \, , \,  A_{00}^{21}=f_{i}^{k} \left(f_{j}^{k}\right){}^* \, , \,  A_{00}^{22}=f_{i}^{k} \left(f_{i}^{k}\right){}^* \, , \, A_{00}^{33}=f_{ij}^{{kl}} \left(f_{ij}^{{kl}}\right){}^*  \,, \nonumber  \\
&& A_{01}^{01}=\left(f_{j}^{m}\right){}^* \, , \,  A_{01}^{02}=\left(f_{i}^{m}\right){}^* \, , \, A_{01}^{13}=f_{j}^{k} \left(f_{ij}^{km}\right){}^* \, , \,  A_{01}^{23}=f_{i}^{k} \left(f_{ij}^{km}\right){}^*  \,, \nonumber   \\
&& A_{02}^{01}=\left(f_{j}^{n}\right){}^*  A_{02}^{02}=\left(f_{i}^{n}\right){}^* \, , \, A_{02}^{13}=f_{j}^{k} \left(f_{ij}^{nk}\right){}^* \, , \, A_{02}^{23}= f_{i}^{k} \left(f_{ij}^{nk}\right){}^*  \,, \nonumber   \\
&& A_{03}^{03}=\left(f_{ij}^{nm}\right){}^*  \,, \nonumber  \\
&& A_{10}^{10}=f_{j}^{m} \, , \,  A_{10}^{20}=f_{i}^{m} \, , \, A_{10}^{31}=f_{ij}^{km} \left(f_{j}^{k}\right){}^* \, , \,  A_{10}^{32}=f_{ij}^{km} \left(f_{i}^{k}\right){}^*  \,, \nonumber  \\
&& A_{11}^{11}=f_{j}^{m} \left(f_{j}^{m}\right){}^* \, , \,  A_{11}^{12}=f_{j}^{m} \left(f_{i}^{m}\right){}^* \, , \,  A_{11}^{21}=f_{i}^{m} \left(f_{j}^{m}\right){}^* \, , \,  A_{11}^{22}=f_{i}^{m} \left(f_{i}^{m}\right){}^* \, , \, A_{11}^{33}=f_{ij}^{km} \left(f_{ij}^{km}\right){}^*  \,, \nonumber  \\
&& A_{12}^{11}=f_{j}^{m} \left(f_{j}^{n}\right){}^* \, , \,  A_{12}^{12}=f_{j}^{m} \left(f_{i}^{n}\right){}^* \, , \, A_{12}^{21}=f_{i}^{m} \left(f_{j}^{n}\right){}^*  A_{12}^{22}=f_{i}^{m} \left(f_{i}^{n}\right){}^* \, , \, A_{12}^{33}=f_{ij}^{km} \left(f_{ij}^{nk}\right){}^*  \,, \nonumber  \\
&& A_{13}^{13}=f_{j}^{m} \left(f_{ij}^{nm}\right){}^* \, , \,  A_{13}^{23}=f_{i}^{m} \left(f_{ij}^{nm}\right){}^*   \,, \nonumber  \\
&& A_{20}^{10}=f_{j}^{n} \, , \,  A_{20}^{20}=f_{i}^{n} \, , \, A_{20}^{31}=f_{ij}^{nk} \left(f_{j}^{k}\right){}^* \, , \,  A_{20}^{32}=f_{ij}^{nk} \left(f_{i}^{k}\right){}^*  \,, \nonumber  \\
&& A_{21}^{11}=f_{j}^{n} \left(f_{j}^{m}\right){}^* \, , \, A_{21}^{12}=f_{j}^{n} \left(f_{i}^{m}\right){}^* \, , \,  A_{21}^{21}=f_{i}^{n} \left(f_{j}^{m}\right){}^* A_{21}^{22}=f_{i}^{n} \left(f_{i}^{m}\right){}^* \, , \, A_{21}^{33}=f_{ij}^{nk} \left(f_{ij}^{km}\right){}^*  \,, \nonumber  \\
&& A_{22}^{11}=f_{j}^{n} \left(f_{j}^{n}\right){}^* \, , \,  A_{22}^{12}=f_{j}^{n} \left(f_{i}^{n}\right){}^* \, , \,  A_{22}^{21}=f_{i}^{n} \left(f_{j}^{n}\right){}^*  A_{22}^{22}=f_{i}^{n} \left(f_{i}^{n}\right){}^* \, , \, A_{22}^{33}=f_{ij}^{nk} \left(f_{ij}^{nk}\right){}^*  \,, \nonumber  \\
&& A_{23}^{13}=f_{j}^{n} \left(f_{ij}^{nm}\right){}^* \, , \, A_{23}^{23}=f_{i}^{n} \left(f_{ij}^{nm}\right){}^*   \,, \nonumber  \\
&& A_{30}^{30}=f_{ij}^{nm} \,, \nonumber \\
&& A_{31}^{31}=f_{ij}^{nm} \left(f_{j}^{m}\right){}^* \, , \,   A_{31}^{32}=f_{ij}^{nm} \left(f_{i}^{m}\right){}^*  \,, \nonumber  \\
&& A_{32}^{31}=f_{ij}^{nm} \left(f_{j}^{n}\right){}^* \, , \,   A_{32}^{32}=f_{ij}^{nm} \left(f_{i}^{n}\right){}^* \,, \nonumber  \\
&& A_{33}^{33}=f_{ij}^{nm} \left(f_{ij}^{nm}\right){}^* \,. \nonumber \\
\end{eqnarray}

\section{Two-qubit entanglement generation} 

\begin{figure}
	\centering
	\includegraphics[width=\linewidth]{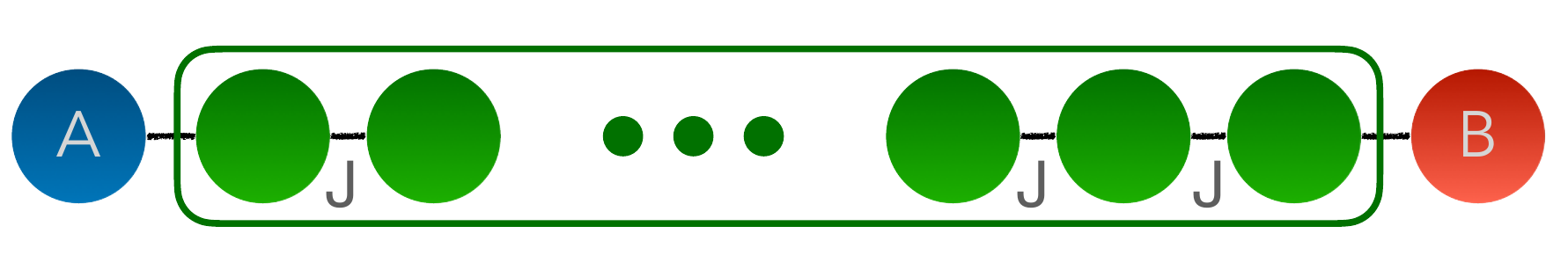}
	\caption{Alice and Bob have access, respectively, to qubits $A$ and $B$, located at the two ends of a spin chain (wire). Entanglement between the two qubit can be generated by exploiting the dynamics of the quantum wire, whose Hamiltonian includes a nearest neighbor interaction term.}
	\label{fig:transfertest}
\end{figure}

In the simplest setting, Alice and Bob aim at generating entanglement between the qubits in their possession, respectively $A$ and $B$, located at some positions in the spin network, by performing local operations on their qubits and exchanging classical communication between them. By definition, LOCC by itself does not allow for neither the increase, nor the generation of entanglement between qubits $A$ and $B$; but the presence of the spin network can give rise to an effective interaction between these qubits, resulting in the possible generation of quantum correlation. In Fig.~\ref{fig:transfertest}, an instance of such an entanglement generation protocol is depicted, where Alice and Bob have access to one spin at each end of a 1D spin chain.
In Ref.~\cite{Wojcik2005}, it has been shown that, by weakly coupling the end qubits to the wire, their state evolves into a Bell state at half of the transfer time of the excitation between the edges. This result has been extended in Ref.~\cite{Wojcik2007} to the generation of a Bell state between two users coupled at arbitrary positions in a spin network, provided control over the local magnetic field is allowed on the sites chosen to be entangled.  Results similar to the weak-coupling scheme can be obtained by strong local magnetic fields on neighboring spins both for one and two-qubit quantum state transfer~\cite{Lorenzo2013a,Lorenzo2015}.  

In Ref.~\cite{Banchi2011}, the authors showed that initialising the system in $\ket{\Psi}_{AB}=\ket{+}_A\ket{+}_B\ket{\psi}_W$, where $\ket{+}=\frac{1}{\sqrt{2}}\left(\ket{0}+\ket{1}\right)$ and $\ket{\psi}_W$ is an arbitrary state with fixed parity of the wire, a maximally entangled state between qubits $A$ and $B$ can be achieved in a ballistic time, provided Alice and Bob can tune the strength of the couplings of their qubits to the wire to an optimal value~\cite{Banchi2011b}. 

\section{Four-qubit entanglement generation}
\label{sec:4q}

\begin{figure}
	\centering
	\includegraphics[width=\linewidth]{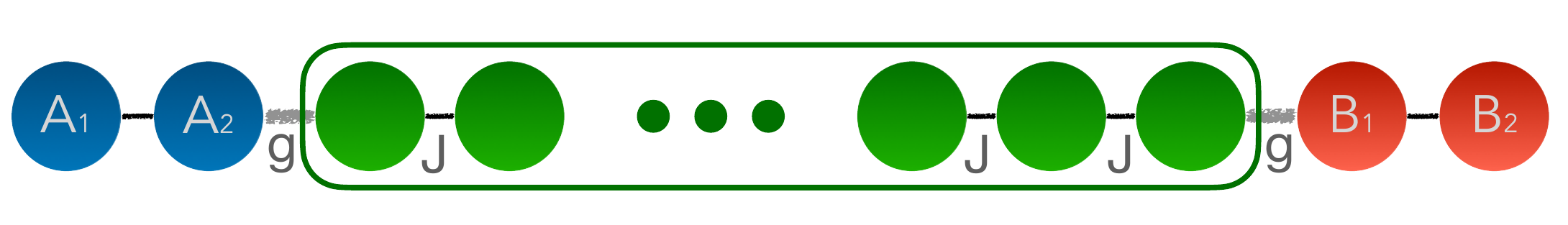}
	\caption{Alice and Bob have access to qubits $A_1, A_2$ and $B_1, B_2$, respectively, located at the two ends of a spin chain (wire), which is capable of generating an entangled state by exploiting its intrinsic dynamics. }
	\label{fig:four_qubit}
\end{figure}

In Sec.~\ref{sec:2q} we have analysed the case where Alice and Bob each have access to one qubit in the spin network. Here, instead, we consider the case where they each have access two qubits. An instance where each pair of qubits is located at an edge of a spin chain is depicted in Fig.~\ref{fig:four_qubit}. A general analysis of the entanglement generated in an arbitrary four-qubit state has not yet been performed, except for pure states \cite{Verstraete2002, Ghahi2016}, due to the complexity of defining entanglement quantifiers for an arbitrary four-qubit mixed state (as a result of the mixed state being in the presence of infinitely many SLOCC classes~\cite{Eltschka2014}).

Restricting only to the $A|B$ partition, in Ref.~\cite{Apollaro2019a} it is shown that starting from $\ket{\Psi}_{AB}=\ket{1100}$ and $\ket{\Psi}_{AB}=\ket{1010}$ (or their mirror-symmetric states $\ket{\Psi}_{AB}=\ket{0011}$ and $\ket{\Psi}_{AB}=\ket{0101}$), a product of two Bell states, i.e., $\ket{\Psi}_{AB}=\ket{\Phi}_{A_1B_2} \otimes \ket{\Phi}_{A_2B_1}$ or $\ket{\Psi}_{AB}=\ket{\Phi}_{A_1B_1} \otimes \ket{\Phi}_{A_2B_2}$, is attained at specific times during the evolution for $g \ll J$. On the other hand, the initial state $\ket{1001}$ (and its mirror-symmetric state $\ket{0110}$) does not generate any entanglement in the $A|B$ partition at any time.

Let us now characterise the type of entanglement these initial states achieve during the evolution of the dynamical map. Starting from the state $\ket{\Psi}_{AB}=\ket{1100}$ results in the evolution
\begin{align}
    \ket{\Psi(t)}_{AB} = \frac{1}{2}\left[ \left(1-\cos\frac{g^2t}{J} \right)\ket{0011} + i\sin\frac{g^2t}{J}\ket{0101} \right.  \,, \nonumber   \\ \left. -i\sin\frac{g^2t}{J}\ket{1010} + \left(1+\cos\frac{g^2t}{J} \right)\ket{1100}\right] \,,
\end{align}
which can be written in a bi-separable form
\begin{align}
    \label{eq:psi1}
    \ket{\Psi(t)}_{AB} = \left|\sin\frac{g^2t}{2J}\right|\left[\left(\ket{01} + -i\cot\frac{g^2t}{2J}\ket{10} \right)_{A_1B_2} \right.  \,, \nonumber  \\ \otimes \left. \left( \ket{01} + i\cot\frac{g^2t}{2J}\ket{10} \right)_{A_2B_1} \right] \,,
\end{align}
implying that the concurrence in each two-qubit state is equal to $C_{A_1B_2} = C_{A_2B_1} = \left| \sin\frac{g^2t}{J} \right|$. This means that a product of two Bell states is created after a time $t = \frac{\pi J}{2g^2}$. It is also noteworthy to consider the four-tangle \cite{Wong2001}, a measure of multipartite entanglement, which is defined for pure states as
\begin{equation}
    \tau_{4}(\ket{\psi}) = \left| \bra{\psi} \sigma_y \otimes \sigma_y \otimes \sigma_y \otimes \sigma_y \ket{\psi^*} \right|^2 \,.
\end{equation}
In this case, the calculation can be carried out explicitly, and we get that $\tau_4(\ket{\Psi(t)}_{AB}) \equiv \tau_{A_1A_2B_1B_2} = \left| \sin\frac{g^2t}{J} \right|^4$. This means that even though the state \eqref{eq:psi1} is biseparable, the four-tangle is non-zero when the two-qubit concurrences are non-zero, implying that the four-tangle is not a measure of exclusive four-way entanglement per se. One can also consider the three-tangle \cite{Coffman2000} of the three-qubit partitions, which contrary to the four-tangle, is exclusively a measure of three-way entanglement. The caveat of this measure is that it is only defined for pure states, while the generalisation to mixed states is described via the convex roof extension. Thus, the three-tangle of a mixed state $\rho$ is given as the average pure state three-tangle minimised over all possible pure state decompositions:
\begin{equation}
    \tau_{3}(\rho) = \min_{\left\{p_i, \ket{\psi_i}\right\}} \sum_{i} p_i \tau_{3}\left(\ket{\psi_i}\right) \,,
\end{equation}
which for a pure state
\begin{equation}
    \tau_{3}(\ket{\psi}) = C_{A(BC)}^2 - C_{AB}^2 - C_{AC}^2 \,,
\end{equation}
and $C_{A(BC)} = \sqrt{2\left(1-\text{Tr}\left(\rho_A^2\right)\right)}$.
If we now consider taking the Eigendecompositions of the partial trace of state \eqref{eq:psi1} with respect to every qubit, we find that the three-tangle of each decomposed pure state for each traced out qubit is equal to zero. This means that we have found a minimal pure state decomposition so that the three-tangle of state \eqref{eq:psi1}, with respect to any partition involving three qubits, is equal to zero, implying there is no three-way entanglement generation between the senders and receivers at any time $t$. This is a consequence of the fact that the initial state does not exhibit any coherence between states having support in the magnetisation sectors with zero and three excitations, and the dynamics are not able to generate any. Clearly, multipartite entanglement distribution protocols are feasible whenever the initial state contains some amount of entanglement, as shown in Ref.~\cite{Apollaro2020} for the case of the three-tangle when the sender's state is GHZ-like.

\begin{figure}[ht]
    %\centering
    %\includegraphics[width=0.33\textwidth]{psi1_chapter_v2.pdf}
    %\includegraphics[width=0.33\textwidth]{psi2_chapter_v1.pdf}
    %\includegraphics[width=0.33\textwidth]{psi3_chapter_v2.pdf}
    \includegraphics[width=1\textwidth]{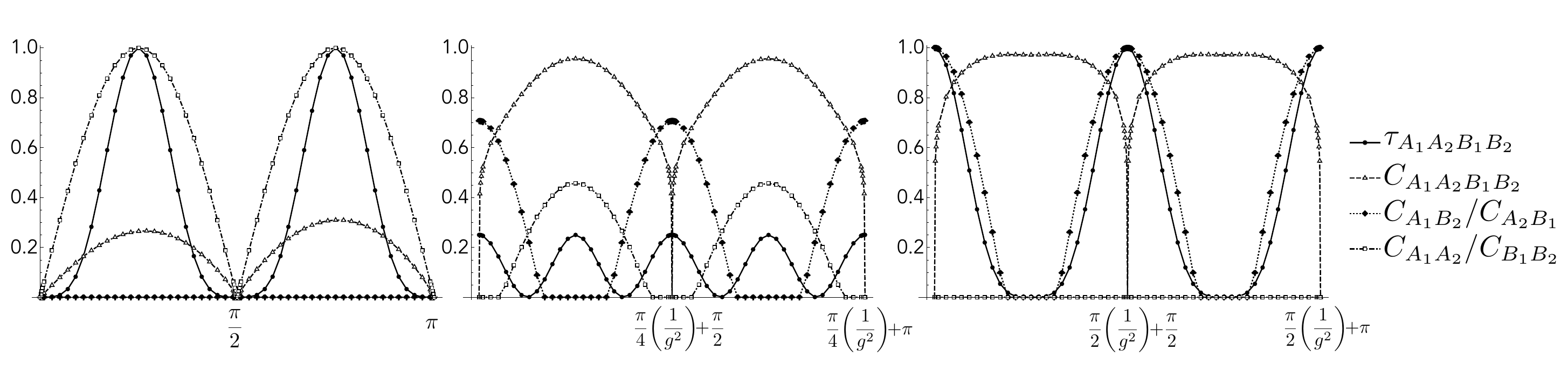}
    \caption{Plot of the two- and the four-qubit concurrences and of the four-tangle for the state \eqref{eq:psi2} with $t$ in units of $J/g^2$. The three panels are at $t=\left[0, \pi\right]$ (left), $t=\left[\frac{\pi}{4g^2}, \frac{\pi}{4g^2}+\pi\right]$  (center), and $t=\left[\frac{\pi}{2g^2}, \frac{\pi}{2g^2}+\pi\right]$ (right), corresponding, respectively, to the beginning, a quarter and an half of the period of the longest time-scale dictated by $g^{-2}$.
    %and $\frac{J^2}{g^2}$ set to unity.
    }
    \label{fig:psi2}
\end{figure}

Let us now move to analyse the evolution of the state generated from $\ket{\Psi}_{AB}=\ket{1010}$. We get that the evolution results in
\begin{align}
\label{eq:psi2}
    \ket{\Psi(t)}_{AB} = \frac{1}{2}\left[ \left(\cos2Jt + \cos\frac{g^2t}{J} \right)\ket{1010} + \left(\cos2Jt - \cos\frac{g^2t}{J} \right)\ket{0101} \right.  \,, \nonumber  \\ \left. -i\sin2Jt\left(\ket{1001} + \ket{0110} \right) -i\sin\frac{g^2t}{J}\left(\ket{1100} - \ket{0011} \right)\right] \,.
\end{align}
To characterise the different types of entanglement in Eq. \eqref{eq:psi2}, we need to look towards a new entanglement measure further to the two-qubit concurrence, and three- and four-tangles. We will specifically use the four-qubit concurrence given in Ref. \cite{Love2007}, which is defined for a pure state as
\begin{align}
    C_{1234} = \left (C_{1(234)}C_{2(134)}C_{3(124)}C_{4(123)}C_{(12)(34)}C_{(13)(24)}C_{(14)(23)} \right)^\frac{1}{7} \,,
\end{align}
where $C_{A(BCD)} = \sqrt{2\left(1-\text{Tr}\left(\rho_A^2\right)\right)}$ and $C_{(AB)(CD)} = \sqrt{\frac{4}{3}\left(1-\text{Tr}\left(\rho_{AB}^2\right)\right)}$. The $n$-partite concurrence is essentially the geometric mean of the concurrence over the set of all possible bipartitions, which is simlar to the GME concurrence \cite{Ma2011}, although the latter is defined as the minimum value of the concurrence over all bipartitions. This inherently implies that the four-qubit concurrence is zero if and only if the four-qubit pure state is separable to some degree. Plotting the four-qubit concurrence $C_{A_1A_2B_1B_2}$ along with the two-qubit concurrences $C_{A_1B_2} (= C_{A_2B_1})$ and $C_{A_1A_2} (= C_{B_1B_2})$, and the four-tangle $\tau_{A_1A_2B_1B_2}$, for $\frac{J^2}{g^2} = 10^4$, we obtain Fig. \ref{fig:psi2}.

The two-qubit concurrences $C_{A_1B_1}$ and $C_{A_2B_2}$ are equal to zero at all times $t$. We make a note that at time $t = \frac{\pi J}{2g^2}$ we acquire a product of two Bell states similar to when we use state \eqref{eq:psi1}. Once again, by taking the Eigendecompositions of the partial trace of state \eqref{eq:psi2} with respect to every qubit, we find that the three-tangle is zero at all times $t$ for every partition consisting of three qubits. Combining this with the fact that at certain times $t$, we have that the only non-zero two-qubit concurrences are $C_{A_1A_2}$ and $C_{B_1B_2}$, and that the four-qubit concurrence $C_{A_1A_2B_1B_2}$ is also non-zero, meaning that the state is fully inseparable, implies that there must be some non-zero value of exclusive four-way entanglement shared between the senders and receivers.

\section{Conclusion}
\label{sec:conc}
In this chapter, we have adopted the general framework of quantum dynamical maps in order to investigate some instances of controlled quantum information dynamics on spin networks. Specifically, we considered those maps emerging from the dynamics of a subset of spin-$\frac{1}{2}$ particles embedded in a larger network, that we divided into sender and receiver parts for convenience. Focusing on the case of a $U(1)$-symmetric Hamiltonian governing the dynamics of the network, we have derived the explicit form of the dynamical map in terms of the excitation transfer amplitude and applied it to review both single-qubit quantum state transfer and two-qubit transfer and corresponding entanglement generation. Finally, we have considered a specific topology of the network where analytical solutions are available for the transfer amplitude and have shown that the quantum map formalism allows the analysis of more complex scenarios such as multi-qubit entanglement generation. It is interesting to note that Ref. \cite{Anticoli2018} provides another means of investigating entanglement generation in quantum state transfer protocols which has not been investigated in this chapter.

The range of applicability of the illustrated approach goes well beyond the cases investigated in this chapter, as it can be easily extended to include classical communication feedback, constraints on the achievable quantum information protocols stemming from the symmetries of the Hamiltonian reflected in the quantum map, and spin networks exposed to external noise.

\section{Acknowledgments}
TJGA ackowledges funding from the UM SEA-EU Research Seed Fund STATED.

\bibliographystyle{unsrt}
\bibliography{springer.bib}

\begin{thebibliography}{10}

\bibitem{Nielsen2010}
Michael~A. Nielsen and Isaac~L. Chuang.
\newblock {\em {Quantum Computation and Quantum Information}}.
\newblock Cambridge University Press, Cambridge, 2010.

\bibitem{Franchini2017}
Fabio Franchini.
\newblock An introduction to integrable techniques for one-dimensional quantum
  systems.
\newblock {\em Lecture Notes in Physics}, 2017.

\bibitem{Orus2019}
Rom{\'{a}}n Or{\'{u}}s.
\newblock {Tensor networks for complex quantum systems}.
\newblock {\em Nature Reviews Physics}, 1(9):538--550, sep 2019.

\bibitem{Society2004}
Ingemar Bengtsson and Karol Zyczkowski.
\newblock {\em {Geometry of Quantum States}}.
\newblock Cambridge University Press, Cambridge, 2017.

\bibitem{Qin2013}
Wei Qin, Chuan Wang, and Gui~Lu Long.
\newblock High-dimensional quantum state transfer through a quantum spin chain.
\newblock {\em Phys. Rev. A}, 87:012339, Jan 2013.

\bibitem{Jaksch2005}
D.~Jaksch and P.~Zoller.
\newblock {The cold atom Hubbard toolbox}.
\newblock {\em Annals of Physics}, 315(1):52--79, jan 2005.

\bibitem{Simon2011}
J.~Simon, W.S. Bakr, Ruichao Ma, M~Eric Tai, P.M. Preiss, and M.~Greiner.
\newblock {Quantum simulation of antiferromagnetic spin chains in an optical
  lattice}.
\newblock {\em Nature}, 472(7343):307--312, 2011.

\bibitem{Bloch2012}
Immanuel Bloch, Jean Dalibard, and Sylvain Nascimb{\`{e}}ne.
\newblock {Quantum simulations with ultracold quantum gases}.
\newblock {\em Nature Physics}, 8(4):267--276, apr 2012.

\bibitem{Labuhn2016b}
Henning Labuhn, Daniel Barredo, Sylvain Ravets, Sylvain
  de~L{\'{e}}s{\'{e}}leuc, Tommaso Macr{\`{i}}, Thierry Lahaye, and Antoine
  Browaeys.
\newblock {Realizing quantum Ising models in tunable two-dimensional arrays of
  single Rydberg atoms}.
\newblock {\em Nature}, 534(7609):667--670, sep 2015.

\bibitem{PhysRevX.8.021069}
Elmer Guardado-Sanchez, Peter~T. Brown, Debayan Mitra, Trithep Devakul,
  David~A. Huse, Peter Schau\ss{}, and Waseem~S. Bakr.
\newblock Probing the quench dynamics of antiferromagnetic correlations in a 2d
  quantum ising spin system.
\newblock {\em Phys. Rev. X}, 8:021069, Jun 2018.

\bibitem{Browaeys2020}
Antoine Browaeys and Thierry Lahaye.
\newblock {Many-body physics with individually controlled Rydberg atoms}.
\newblock {\em Nature Physics}, 16(2):132--142, 2020.

\bibitem{Pitsios2017a}
Ioannis Pitsios, Leonardo Banchi, Adil~S Rab, Marco Bentivegna, Debora Caprara,
  Andrea Crespi, Nicol{\`{o}} Spagnolo, Sougato Bose, Paolo Mataloni, Roberto
  Osellame, and Fabio Sciarrino.
\newblock {Photonic simulation of entanglement growth and engineering after a
  spin chain quench}.
\newblock {\em Nature Communications}, 8(1):1569, dec 2017.

\bibitem{Porras2004}
D.~Porras and J.~I. Cirac.
\newblock {Effective Quantum Spin Systems with Trapped Ions}.
\newblock {\em Physical Review Letters}, 92(20):1--4, may 2004.

\bibitem{Friedenauer2008}
A.~Friedenauer, H.~Schmitz, J.~T. Glueckert, D.~Porras, and T.~Schaetz.
\newblock {Simulating a quantum magnet with trapped ions}.
\newblock {\em Nature Physics}, 4(10):757--761, jul 2008.

\bibitem{Monroe2021}
C.~Monroe, W.~C. Campbell, L.~M. Duan, Z.~X. Gong, A.~V. Gorshkov, P.~W. Hess,
  R.~Islam, K.~Kim, N.~M. Linke, G.~Pagano, P.~Richerme, C.~Senko, and N.~Y.
  Yao.
\newblock {Programmable quantum simulations of spin systems with trapped ions}.
\newblock {\em Reviews of Modern Physics}, 93(2):25001, 2021.

\bibitem{Kay2008}
A.~Kay and D.~G. Angelakis.
\newblock {Reproducing spin lattice models in strongly coupled atom-cavity
  systems}.
\newblock {\em Epl}, 84(2), 2008.

\bibitem{Noh2017}
Changsuk Noh and Dimitris~G. Angelakis.
\newblock {Quantum simulations and many-body physics with light}.
\newblock {\em Reports on Progress in Physics}, 80(1), 2017.

\bibitem{Heras2014}
U.~Las Heras, A.~Mezzacapo, L.~Lamata, S.~Filipp, A.~Wallraff, and E.~Solano.
\newblock {Digital quantum simulation of spin systems in superconducting
  circuits}.
\newblock {\em Physical Review Letters}, 112(20):1--5, 2014.

\bibitem{Vepsalainen2020}
Antti Veps{\"{a}}l{\"{a}}inen and Gheorghe~Sorin Paraoanu.
\newblock {Simulating Spin Chains Using a Superconducting Circuit: Gauge
  Invariance, Superadiabatic Transport, and Broken Time-Reversal Symmetry}.
\newblock {\em Advanced Quantum Technologies}, 3(4):1--12, 2020.

\bibitem{Gisin2007}
Nicolas Gisin and Rob Thew.
\newblock {Quantum communication}.
\newblock {\em Nature Photonics}, 1(3):165--171, 2007.

\bibitem{Pirandola2017}
Stefano Pirandola, Riccardo Laurenza, Carlo Ottaviani, and Leonardo Banchi.
\newblock {Fundamental limits of repeaterless quantum communications}.
\newblock {\em Nature Communications}, 8, 2017.

\bibitem{book:19074}
K.~Kraus, A.~Böhm, J.~D. Dollard, and W.~H. Wootters.
\newblock {\em States, Effects, and Operations Fundamental Notions of Quantum
  Theory}.
\newblock LNP0190. Springer, 1983.

\bibitem{Stinespring}
W.~Forrest Stinespring.
\newblock Positive functions on c*-algebras.
\newblock {\em Proceedings of the American Mathematical Society},
  6(2):211--216, 1955.

\bibitem{Kraus}
K~Kraus.
\newblock General state changes in quantum theory.
\newblock {\em Annals of Physics}, 64(2):311--335, 1971.

\bibitem{Zyczkowskibook}
K.~{\.Z}yczkowski, I.~Bengtsson, and an~O'Reilly Media~Company Safari.
\newblock {\em Geometry of Quantum States, 2nd Edition}.
\newblock Cambridge University Press, 2017.

\bibitem{Verma2017b}
Harshit Verma, L.~Chotorlishvili, J.~Berakdar, and Sunil~K. Mishra.
\newblock {Qubit(s) transfer in helical spin chains}.
\newblock {\em EPL (Europhysics Letters)}, 119(3):30001, aug 2017.

\bibitem{Campbell2013d}
S~Campbell, L~Mazzola, G~{De Chiara}, T~J~G Apollaro, F~Plastina, Th~Busch, and
  M~Paternostro.
\newblock {Global quantum correlations in finite-size spin chains}.
\newblock {\em New Journal of Physics}, 15(4):043033, apr 2013.

\bibitem{PhysRevA.93.012343}
Xining Chen, Robert Mereau, and David~L Feder.
\newblock {Asymptotically perfect efficient quantum state transfer across
  uniform chains with two impurities}.
\newblock {\em Phys. Rev. A}, 93(1):12343, 2016.

\bibitem{Bose2003}
Sougato Bose.
\newblock {Quantum Communication through an Unmodulated Spin Chain}.
\newblock {\em Physical Review Letters}, 91(20):1--4, nov 2003.

\bibitem{Bayat2011}
Abolfazl Bayat, Leonardo Banchi, Sougato Bose, and Paola Verrucchi.
\newblock {Initializing an unmodulated spin chain to operate as a high-quality
  quantum data bus}.
\newblock {\em Physical Review A}, 83(6):1--9, jun 2011.

\bibitem{Amico2004b}
Luigi Amico, Andreas Osterloh, Francesco Plastina, Rosario Fazio, and
  G.~{Massimo Palma}.
\newblock {Dynamics of entanglement in one-dimensional spin systems}.
\newblock {\em Physical Review A}, 69(2):022304, feb 2004.

\bibitem{Fubini2006a}
A.~Fubini, T.~Roscilde, V.~Tognetti, M.~Tusa, and P.~Verrucchi.
\newblock {Reading entanglement in terms of spin configurations in quantum
  magnets}.
\newblock {\em The European Physical Journal D}, 38(3):563--570, apr 2006.

\bibitem{Bose2007}
Sougato Bose.
\newblock {Quantum communication through spin chain dynamics: an introductory
  overview}.
\newblock {\em Contemporary Physics}, 48(1):13--30, jan 2007.

\bibitem{Burgarth2005}
Daniel Burgarth and Sougato Bose.
\newblock {Perfect quantum state transfer with randomly coupled quantum
  chains}.
\newblock {\em New Journal of Physics}, 7:135--135, may 2005.

\bibitem{Apollaro2010a}
T~J~G Apollaro, A~Cuccoli, C~{Di Franco}, M~Paternostro, F~Plastina, and
  P~Verrucchi.
\newblock {Manipulating and protecting entanglement by means of spin
  environments}.
\newblock {\em New Journal of Physics}, 12(8):083046, aug 2010.

\bibitem{Apollaro2015}
T~J~G Apollaro, S.~Lorenzo, A.~Sindona, S.~Paganelli, G~L Giorgi, and
  F.~Plastina.
\newblock {Many-qubit quantum state transfer via spin chains}.
\newblock {\em Physica Scripta}, T165(T165):014036, oct 2015.

\bibitem{Wojcik2005}
Antoni W{\'{o}}jcik, Tomasz {\L}uczak, Pawe{\l} Kurzy{\'{n}}ski, Andrzej
  Grudka, Tomasz Gdala, and Ma{\l}gorzata Bednarska.
\newblock {Unmodulated spin chains as universal quantum wires}.
\newblock {\em Physical Review A}, 72(3):034303, sep 2005.

\bibitem{Wojcik2007}
Antoni W{\'{o}}jcik, Tomasz {\L}uczak, Pawe{\l} Kurzy{\'{n}}ski, Andrzej
  Grudka, Tomasz Gdala, and Ma{\l}gorzata Bednarska.
\newblock {Multiuser quantum communication networks}.
\newblock {\em Physical Review A}, 75(2):022330, feb 2007.

\bibitem{Lorenzo2013a}
Salvatore Lorenzo, Tony J.~G. Apollaro, Antonello Sindona, and Francesco
  Plastina.
\newblock {Quantum-state transfer via resonant tunneling through
  local-field-induced barriers}.
\newblock {\em Phys. Rev. A}, 87(0):042313, apr 2013.

\bibitem{Lorenzo2015}
S.~Lorenzo, T.~J.~G. Apollaro, S.~Paganelli, G.~M. Palma, and F.~Plastina.
\newblock Transfer of arbitrary two-qubit states via a spin chain.
\newblock {\em Phys. Rev. A}, 91:042321, Apr 2015.

\bibitem{Banchi2011}
Leonardo Banchi, Abolfazl Bayat, Paola Verrucchi, and Sougato Bose.
\newblock {Nonperturbative Entangling Gates between Distant Qubits Using
  Uniform Cold Atom Chains}.
\newblock {\em Physical Review Letters}, 106(14):140501, apr 2011.

\bibitem{Banchi2011b}
Leonardo Banchi, Tony J.~G. Apollaro, and Alessandro Cuccoli.
\newblock {Long quantum channels for high-quality entanglement transfer}.
\newblock {\em New Journal of Physics}, 123006, 2011.

\bibitem{Verstraete2002}
F.~Verstraete, J.~Dehaene, B.~{De Moor}, and H.~Verschelde.
\newblock {Four qubits can be entangled in nine different ways}.
\newblock {\em Phys. Rev. A - At. Mol. Opt. Phys.}, 65(5):521121--521125, 2002.

\bibitem{Ghahi2016}
Masoud~Gharahi Ghahi and Seyed~Javad Akhtarshenas.
\newblock {Entangled graphs: A classification of four-qubit entanglement}.
\newblock {\em Eur. Phys. J. D}, 70(3):1--6, 2016.

\bibitem{Eltschka2014}
Christopher Eltschka and Jens Siewert.
\newblock {Quantifying entanglement resources}.
\newblock {\em Journal of Physics A: Mathematical and Theoretical}, 47(42),
  2014.

\bibitem{Apollaro2019a}
Tony J~G Apollaro, Guilherme M~A Almeida, Salvatore Lorenzo, Alessandro
  Ferraro, and Simone Paganelli.
\newblock {Spin chains for two-qubit teleportation}.
\newblock {\em Physical Review A}, 100(5):052308, nov 2019.

\bibitem{Wong2001}
Alexander Wong and Nelson Christensen.
\newblock Potential multiparticle entanglement measure.
\newblock {\em Phys. Rev. A}, 63:044301, Mar 2001.

\bibitem{Coffman2000}
Valerie Coffman, Joydip Kundu, and William~K. Wootters.
\newblock {Distributed entanglement}.
\newblock {\em Phys. Rev. A}, 61(5):5, 2000.

\bibitem{Apollaro2020}
Tony~J.G. Apollaro, Claudio Sanavio, Wayne~Jordan Chetcuti, and Salvatore
  Lorenzo.
\newblock {Multipartite entanglement transfer in spin chains}.
\newblock {\em Physics Letters A}, 384(15):126306, may 2020.

\bibitem{Love2007}
Peter~J. Love, Alec~Maassen {Van Den Brink}, A.~Yu Smirnov, M.~H.S. Amin,
  M.~Grajcar, E.~Il'ichev, A.~Izmalkov, and A.~M. Zagoskin.
\newblock {A characterization of global entanglement}.
\newblock {\em Quantum Inf. Process.}, 6(3):187--195, 2007.

\bibitem{Ma2011}
Zhi~Hao Ma, Zhi~Hua Chen, Jing~Ling Chen, Christoph Spengler, Andreas Gabriel,
  and Marcus Huber.
\newblock {Measure of genuine multipartite entanglement with computable lower
  bounds}.
\newblock {\em Phys. Rev. A}, 83(6):1--5, 2011.

\bibitem{Anticoli2018}
Linda Anticoli and Masoud~Gharahi Ghahi.
\newblock Modeling tripartite entanglement in quantum protocols using evolving
  entangled hypergraphs.
\newblock {\em International Journal of Quantum Information}, 16(07):1850055,
  2018.

\end{thebibliography}

\end{document}